\documentclass[aps,prb,floatfix,twocolumn]{revtex4} 
\usepackage{graphicx}
\usepackage{amsmath,amsfonts,amssymb}
\usepackage{color}
\newcommand{\nairo}{Na$_2$IrO$_3$}

\newcommand{\aliiro}{$\alpha$-Li$_2$IrO$_3$}
\newcommand{\bliiro}{$\beta$-Li$_2$IrO$_3$}
\newcommand{\gliiro}{$\gamma$-Li$_2$IrO$_3$}

\begin{document}
\title{Electronic excitations in {\gliiro}}
\date{\today}
\begin{abstract}
  We investigate the electronic properties of the three-dimensional
  stripyhoneycomb {\gliiro} via relativistic density functional theory
  calculations as well as exact diagonalization of finite clusters and
  explore the details of the optical conductivity.  Our analysis of
  this quantity reveals the microscopic origin of the experimentally
  observed (i) optical transitions and (ii) anisotropic
  behavior along the various polarization directions. 
  In particular we find
that the optical
excitations are overall dominated by transitions between $j_\text{eff}$ = 1/2
and 3/2 states and the weight of transitions between $j_\text{eff}$ = 1/2
states at low frequencies can be correlated to deviations from a pure Kitaev description.  We furthermore reanalyze within this
approach the electronic excitations in the
  known two-dimensional honeycomb systems $\alpha$-Li$_2$IrO$_3$ and
  Na$_2$IrO$_3$ and discuss the results in comparison to {\gliiro}.
\end{abstract}

\author{Ying Li}
\affiliation{Institut f\"ur Theoretische Physik, Goethe-Universit\"at Frankfurt,
Max-von-Laue-Strasse 1, 60438 Frankfurt am Main, Germany}
\author{Stephen M. Winter}
\affiliation{Institut f\"ur Theoretische Physik, Goethe-Universit\"at Frankfurt,
Max-von-Laue-Strasse 1, 60438 Frankfurt am Main, Germany}
\author{Harald O. Jeschke}
\affiliation{Institut f\"ur Theoretische Physik, Goethe-Universit\"at Frankfurt,
Max-von-Laue-Strasse 1, 60438 Frankfurt am Main, Germany}
\author{Roser Valent{\'\i}}
\affiliation{Institut f\"ur Theoretische Physik, Goethe-Universit\"at Frankfurt,
Max-von-Laue-Strasse 1, 60438 Frankfurt am Main, Germany}

\maketitle
\section{Introduction}

The two-dimensional honeycomb iridates {\nairo} and {\aliiro} have
been suggested as candidate materials for the realization of
bond-dependent anisotropic interactions as described by the Kitaev
model~\cite{Kitaev2006}. The appropriate description of the electronic
structure of these materials is currently being discussed. In the
limit of strong spin-orbit coupling (SOC) and electron-electron interactions, the low-energy degrees of freedom are predicted to be localized spin-orbital doublet $j_{\rm eff}=1/2$ states\cite{Jackeli2009,
  Chaloupka2010, Imada2014, Katukuri2014, Rau2014,
  Reuther2014}. These localized moments are thought to persist despite relatively weak correlations in the $5d$ Ir orbitals due to an effective bandwidth reduction via SOC. That is, once SOC is included, the highest occupied $j_{\rm eff}=1/2$ bands become very narrow, enhancing the role of correlations. A complementary perspective was also given from the limit of weak correlations. In this case, the electronic properties
of these systems can be described in terms of a recently proposed
quasimolecular orbital (QMO) basis\cite{Mazin2012, Foyevtsova2013,
  Li2015}. When SOC is included in this picture, a (pseudo)gap was found at the Fermi energy for {\nairo}, suggesting the material is relatively close to a band insulating state in the weak correlation limit.\cite{Mazin2012} Thus even weak correlations may be sufficient to induce an insulating state. Overall, a correct understanding of the electronic structure is important for evaluating the relevance of localized spin-Hamiltonians such as the (extended) Heisenberg-Kitaev models currently under discussion for these materials. It is generally agreed that long-range second and/or third neighbour interactions are required to understand the magnetism in the 2D honeycomb {\nairo} and {\aliiro}~\cite{Winter2016}, suggesting relatively delocalized moments.

\begin{figure}
\includegraphics[angle=0,width=0.95\linewidth]{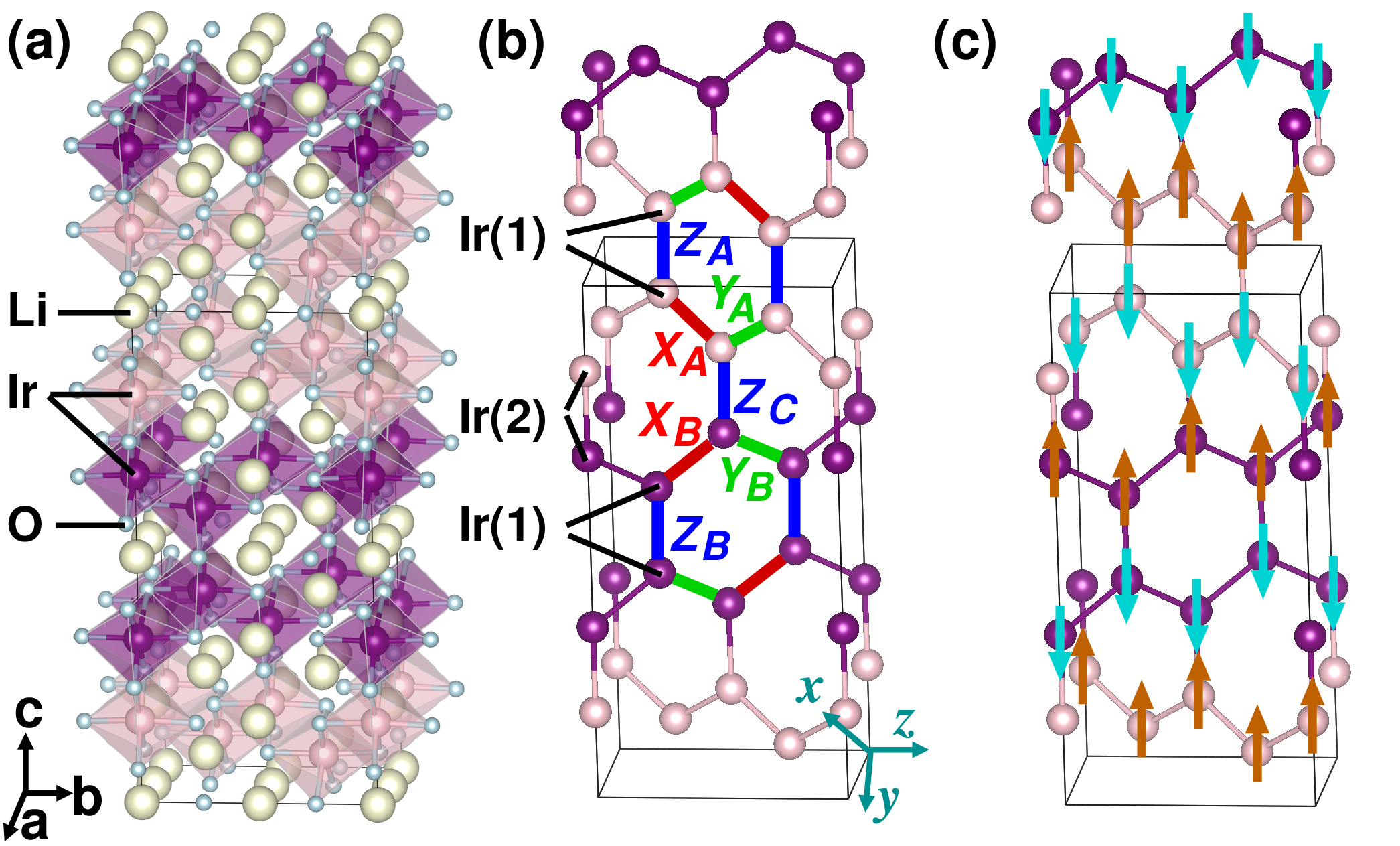}
\caption{(Color online)  (a) Crystal structure of stripyhoneycomb {\gliiro}\cite{Biffin2014gamma, Modic2014}. Honeycomb rows
alternate in orientation along $c$. The black axis $a$, $b$, and $c$ are the vectors of the unit
cell. (b) Crystal structure showing only Ir atoms. The red, green and blue bonds show the seven different types of bonds $X_A$, $X_B$, $Y_A$, $Y_B$, $Z_A$, $Z_B$, $Z_C$. $x$, $y$, $z$ are the cartesian coordinates for the $d$ orbitals. (c) Zigzag magnetic configuration used in our GGA+SO+U calculations.}
\label{fig:structure}
\end{figure}

Three-dimensional generalizations of the honeycomb lattices were also
recently synthesized; the hyperhoneycomb {\bliiro}~\cite{Takayama2015,
  Biffin2014beta} and the stripyhoneycomb {\gliiro}~\cite{Modic2014,
  Biffin2014gamma} (Fig. \ref{fig:structure}).  These materials are expected to display 3D Kitaev
physics and to potentially support quantum spin liquid states
analogous to the 2D case~\cite{Mandal2009, Kimchi2014,
  Lee2014}. Resonant magnetic x-ray diffraction experiments found that
{\gliiro} hosts, at low temperatures, a non-coplanar counter-rotating
long range spiral magnetic order with incommensurate ordering
wavevector ${\bf q} = (0.57,0,0)$ along the orthorhombic $a$-axis~\cite{Biffin2014beta, Biffin2014gamma}. Various investigations of
the combined Kitaev-Heisenberg spin Hamiltonian suggest that Kitaev
interactions must dominate over the Heisenberg terms in order to
produce the observed complex spin spirals~\cite{Kimchi2014,
  Kimchi2015, Lee2014, Reuther2014,Lee2016, Biffin2014gamma}, although
long-range antisymmetric interactions cannot be
ignored~\cite{Winter2016}.

In order to gain microscopic insight on the electronic properties of
{\gliiro} in comparison to its 2D counterparts, we consider the electronic structure and optical conductivity of each material within density functional theory (DFT) and the exact diagonalization (ED) method. Optical conductivity
measurements for {\gliiro}~\cite{Hinton2015} show anisotropic
behavior between polarizations along the $a$ and $b$ axes, but
both polarizations show a broad peak structure at 1.5~eV,  similar to that of {\nairo}.  However, the observed optical conductivity was significantly reduced in magnitude for $\gamma$-Li$_2$IrO$_3$ compared to {\nairo}. This difference was initially attributed to the inherently 3D versus 2D structure rather than the
replacement of Na by Li \cite{Hinton2015}. This issue is addressed in Section \ref{sec-D}. The remaining paper is organized as follows. In section \ref{sec-1}, we discuss the electronic structure of $\gamma$-Li$_2$IrO$_3$ from the perspective of both DFT calculations and exact diagonalization of small clusters. In section \ref{sec-D}, we relate the electronic structure to the optical conductivity, including detailed discussion of the differences between DFT and ED results. Finally, in section \ref{sec-3} we compare the results for $\gamma$-Li$_2$IrO$_3$ to the 2D honeycomb lattice analogues Na$_2$IrO$_3$ and $\alpha$-Li$_2$IrO$_3$. In particular, in this last section, we present results based on the newly available single-crystal structure of $\alpha$-Li$_2$IrO$_3$ \cite{Freund2016}.

\section{Electronic properties of {\protect\gliiro} \label{sec-1}}
\subsection{ Crystal Structure}

The experimental structure of {\gliiro}~\cite{Biffin2014gamma,
  Modic2014} (see Fig.~\ref{fig:structure} (a)-(b)) has two hexagonal
chains oriented in the directions $a$ $\pm$ $b$ linked
along the $c$-direction.  There are three kinds of $Z$ bonds in
{\gliiro}: the $Z_C$ bond bridges two chains of hexagons while the
$Z_A$ and $Z_B$ bonds complete each Ir hexagon in the layered
structure. The cartesian coordinates $x$, $y$, $z$ for the orbitals
are displayed in Fig.~\ref{fig:structure} (b). The unit cell has two
nonequivalent Ir atoms and a total of eight Ir: Ir(1) atoms are linked
by $Z_A$ and $Z_B$ bonds, while Ir(2) atoms are linked by $Z_C$
bonds. $X_A$, $Y_A$, $X_B$ and $Y_B$ link Ir(1) and Ir(2)
sites. Details of the crystal structure are given in Table I.
\begin{table}[t]
\caption {Nearest neighbour distances  (in {\AA})  and Ir-O-Ir angles for the different bond types, determined in the experimental  {\gliiro} structure
    (see Fig.~\ref{fig:structure} (b) for bond notation).}
\label{table:struct}
\centering
\def\arraystretch{1.1}
\begin{ruledtabular}
\begin{tabular}{rrrrr}
 {\gliiro}                 & $X_A$, $Y_A$   & $Y_B$, $X_B$ &            $Z_A$, $Z_B$   &  $Z_C$ \\
\hline                                                                                  
 Ir-Ir distance   &\multicolumn{2}{c}{2.976}                 &2.982       & 2.96  \\
 Ir-O1 distance   &\multicolumn{2}{c}{1.99,2.14}         &2.10       & 1.97  \\
 Ir-O2 distance   &\multicolumn{2}{c}{2.01,2.01}         &2.10       & 1.97  \\
 Ir-O1-Ir angle   &\multicolumn{2}{c}{92.00$^\circ$}                 &90.37$^\circ$       & 97.40$^\circ$  \\
 Ir-O2-Ir angle   &\multicolumn{2}{c}{95.52$^\circ$}                 &90.37$^\circ$      & 97.40$^\circ$  \\
\end{tabular}
\end{ruledtabular}
\end{table}

\subsection{Density Functional Theory Calculations}
We performed linearized augmented plane-wave (LAPW) calculations~\cite{Blaha2001} with the generalized gradient
approximation (GGA)\cite{Perdew1996}. We chose the basis-size
controlling parameter $RK_{\rm max} = 8$ and a mesh of 432 {\bf k}
points in the first Brillouin zone (FBZ) of the primitive unit
cell. Relativistic effects were taken into account within the second
variational approximation. A $U_{\rm eff}=2.4 $~eV as implemented in
GGA+SO+U\cite{Anisimov1993} was employed in order to keep consistency
with previous calculations~\cite{Li2015}.
The hopping parameters between Ir $5d$ orbitals in {\gliiro} were
computed via the Wannier function projection
method~\cite{Aichorn2009,Foyevtsova2013,Ferber2014,Winter2016} and we
employed the optics code package~\cite{Draxl2006} within LAPW to
calculate the optical conductivity.  The density of states and optical
properties were computed with 10 $\times$ 10 $\times$ 10 {\bf k}
points in the full Brillouin zone while the hopping parameters were
evaluated using 12 $\times$ 12 $\times$ 12 {\bf k} points. 
%

\begin{figure}
  \includegraphics[angle=0,width=0.95\linewidth]{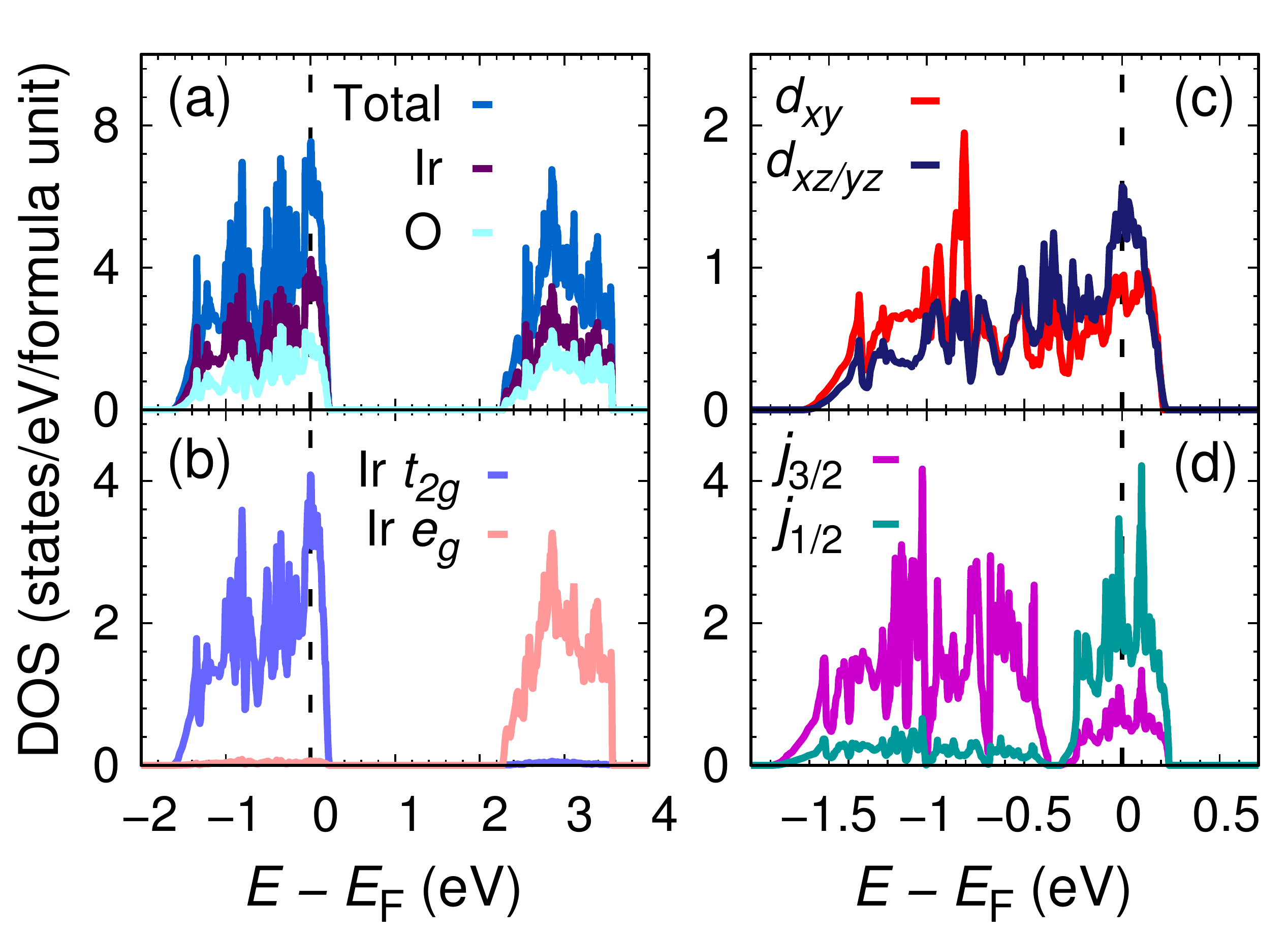}
  \caption{(Color online) Density of states (DOS) for {\gliiro} in the
    nonmagnetic configuration obtained (a-c) within GGA and (d)
    GGA+SO.}
\label{ggados}
\end{figure}

 The non-relativistic GGA density of states (DOS) for
 $\gamma$-Li$_2$IrO$_3$ is displayed in Fig.~\ref{ggados} (a) -
 (c). The Iridium $5 d$ states are split into $e_g$ (2.2 eV to 3.6 eV)
 and $t_{2g}$ (-1.6 eV to 0.2 eV) states (Fig.~\ref{ggados} (b)) due
 to the octahedral crystal field of IrO$_6$ with the Fermi level lying
 within the $t_{2g}$ manifold.  The $t_{2g}$ band is further slightly
 split into lower $d_{xy}$ and higher $d_{xz}$, $d_{yz}$
 (Fig.~\ref{ggados} (c)), arising from an additional weak trigonal
 field. By using the projection method described in
Ref.~\onlinecite{Foyevtsova2013}, we obtained the hopping parameters
from the GGA bandstructure.  Table~\ref{table:cfs} displays the
crystal field splitting compared with {\nairo}. Full hopping integral
tables are given in Appendix A. In terms of the $t_{2g}$ $d$-orbital basis:
\begin{align}
\vec{\mathbf{c}}_i^\dagger = \left(c^\dagger_{i,yz,\uparrow} \  c^\dagger_{i,yz,\downarrow} \ c^\dagger_{i,xz,\uparrow} \  c^\dagger_{i,xz,\downarrow} \ c^\dagger_{i,xy,\uparrow} \  c^\dagger_{i,xy,\downarrow}\right) 
\end{align}
the crystal field terms can be written:
\begin{align}
\mathcal{H}_{\rm CF}= - \sum_i \vec{\mathbf{c}}_{i}^\dagger\left\{\mathbf{E}_i\otimes \mathbb{I}_{2\times 2}\right\}\vec{\mathbf{c}}_i
\end{align}
where $\mathbb{I}_{2\times 2}$ is the $2 \times 2$ identity matrix (for the spin variables); the crystal field tensor $\mathbf{E}_i$ is constrained by local 2-fold symmetry at each Ir site to be:
\begin{align}
\mathbf{E}_i = \left(\begin{array}{ccc} 0&\Delta_{1}&\Delta_{2} \\ \Delta_{1}&0&\Delta_{2} \\ \Delta_{2} & \Delta_{2} & \Delta_{3}\end{array} \right)
\end{align}
The $t_{2g}$ crystal fields
$\Delta_1$, $\Delta_2$ denote the on-site hopping between $d_{xz}$ and
$d_{yz}$ orbitals, and between $d_{xy}$ and $d_{yz/xz}$ orbitals,
respectively (Table~\ref{table:cfs}). $\Delta_3$ is the on-site energy
of $d_{xy}$ minus that of $d_{yz/xz}$~\cite{Winter2016}. $\Delta_3$ is
-213.5 meV for Ir(1) and -110.9 meV for Ir(2) (see
Fig.~\ref{fig:structure}), which is much larger in magnitude than in
{\nairo} (\mbox{-27.2}~meV)~\cite{Winter2016}. This means that in the 3D
{\gliiro} structure, the $t_{2g}$ crystal field is of the same order
of magnitude as the spin-orbit coupling $\lambda\sim 400$ meV and this likely has
significant effects on the local magnetic interactions.
\begin{table}[t]
  \caption {Crystal field splitting compared with {\nairo}. 
    The $t_{2g}$ crystal fields $\Delta_1$, $\Delta_2$ denote, 
    respectively, the onsite hopping between
    $d_{xz}$ and $d_{yz}$ orbitals, $d_{xy}$ and $d_{yz/xz}$ orbitals. 
    $\Delta_3$ is the on-site energy of $d_{xy}$ minus 
    $d_{yz/xz}$~\cite{Winter2016}.}
\label{table:cfs}
\centering
\def\arraystretch{1.1}
\begin{ruledtabular}
\begin{tabular}{cccc}
Crystal field    &{\nairo} \cite{Winter2016} & \multicolumn{2}{c}{\gliiro}  \\
Paramater &&Ir(1) &  Ir(2) \\
\hline
 $\Delta_1$ &-22.9 &  -24.4  &  -29.9\\
 $|\Delta_2|$ &27.6 &  4.2   & 37.4\\
 $\Delta_3$ &-27.2  &  -213.5 & -110.9\\
\end{tabular}
\end{ruledtabular}
\end{table}


Table~\ref{table:hop} shows the nearest neighbour hopping parameters
where $t_{1\|}$, $t_{1O}$ and $t_{1\sigma}$ are defined as in
Ref.~\onlinecite{Foyevtsova2013} (labelled $t_1$, $t_2$, and $t_3$ in Ref.~\onlinecite{Rau2014, Winter2016}); $t_{1O}$ ($t_2$) refers to effective Ir$-$Ir hopping through the bridging oxygens, $t_{1\sigma}$ ($t_3$) and $t_{1\|}$ ($t_1$) refer to $\sigma$- and $\delta$-type direct metal-metal hopping, respectively. A full table of hopping integrals in the $t_{2g}$ basis are given in the Appendix. There are three
significant differences in the nearest neighbour hoppings of the 3D
$\gamma$-Li$_2$IrO$_3$ (see Table~\ref{table:hop}) when compared with
{\nairo}: i) the direct metal-metal hopping $t_3$ ($d_{xy}
\rightarrow d_{xy}$) along the $Z_A$ and $Z_B$ bonds
(Fig.~\ref{fig:structure} (b)) is larger than the oxygen-assisted
hopping $t_2$ ($d_{xz} \rightarrow d_{yz}$, $d_{yz} \rightarrow
d_{xz}$) due to the nearly $90^\circ$ Ir-O-Ir angle
(Table~\ref{table:struct}). ii) the $t_2$ in the $X_A$ ($Y_A$), $X_B$
($Y_B$) bonds have opposite signs, as a result of different local
environments (see Appendix). The different sign arises because such bonds are related to one another by crystallographic 2-fold rotations. Finally, iii) the absence of inversion symmetry
for the majority of nearest neighbour bonds allows for some asymmetry
in the $t_2$ hopping, e.g. for the $X_A$ bond, $d_{xy} \rightarrow
d_{xz}$ and $d_{xz} \rightarrow d_{xy}$ hoppings are unequal. For this
reason, a finite Dzyaloshinskii-Moriya (DM) interaction is both
allowed and expected to appear for the majority of first-neighbour
bonds: $X_A$, $X_B$, $Y_A$, $Y_B$, and $Z_C$. This result is in
contrast to {\nairo} and {\aliiro}, for which a weaker DM interaction only exists
for the second nearest neighbour bonds \cite{Winter2016}. Since these
antisymmetric interactions are likely to strongly stabilize the
observed incommensurate magnetic order~\cite{Biffin2014gamma}, one may
question the completeness of previous interaction models for {\gliiro}
including only symmetric exchange interactions
\cite{Kimchi2014,Kimchi2015}.
\begin{table}[t]
  \caption{Nearest neighbour hopping integrals in meV between 
    Ir $t_{2g}$ orbitals for the experimental  {\gliiro} structure
    (see Fig.~\ref{fig:structure} (b) for bond notation). The 
    labels $t_{1\|}$, $t_{1O}$ and $t_{1\sigma}$ are the same as in 
    Ref.~\onlinecite{Foyevtsova2013}, and the notations $t_1$, 
    $t_2$ and $t_3$ are given in Ref.~\onlinecite{Rau2014, Winter2016}.}
\label{table:hop}
\centering
\def\arraystretch{1.1}
\begin{ruledtabular}
\begin{tabular}{rrrrr}
 {\gliiro}                 & $X_A$, $Y_A$   & $Y_B$, $X_B$ &            $Z_A$, $Z_B$   &  $Z_C$   \\
\hline                                                                                   
$t_{1\|}$ ($t_1$)  &91.4              &91.4                    &91.8           &77.4     \\
                   &69.2              &69.2                    &91.8           &77.4     \\
$t_{1O}$  ($t_2$)  &-262.5            &262.5                   &132.8          &294.1    \\
                   &-240.5            &240.5                   &132.8          &294.1    \\
$t_{1\sigma}$ ($t_3$)&-168.3          &-168.3                  &-319.7         &-17.1    \\
\end{tabular}
\end{ruledtabular}
\end{table}

Unlike the 2D {\nairo}, the 3D $\gamma$-Li$_2$IrO$_3$
does not allow a clear description of the DFT electronic structure in
terms of QMOs. Indeed, there is no pseudogap at the Fermi energy at the GGA+SO level (Fig. \ref{ggados}(d)), in contrast with {\nairo}. As in the $P\,3_112$ structure of
$\alpha$-RuCl$_3$~\cite{Johnson2015}, the oxygen assisted hopping
$t_{1O}$, which is crucial for the formation of the QMOs, is smaller
than $t_{1\sigma}$ \cite{Foyevtsova2013,Johnson2015}. In addition,
since not all local Ir $5d$ orbitals can be attributed to a single
hexagon, the QMO basis is incomplete.  We therefore choose to work
with the $j_{\rm eff}$ basis.  Fig.~\ref{ggados}~(d) shows the
projection of the GGA+SO DOS onto the $j_{\rm eff}$ basis. At the
Fermi level, the DOS is dominantly $j_{\rm eff}=1/2$ with a small
contribution from $j_{\rm eff}=3/2$. 

\begin{figure}
\includegraphics[angle=0,width=0.95\linewidth]{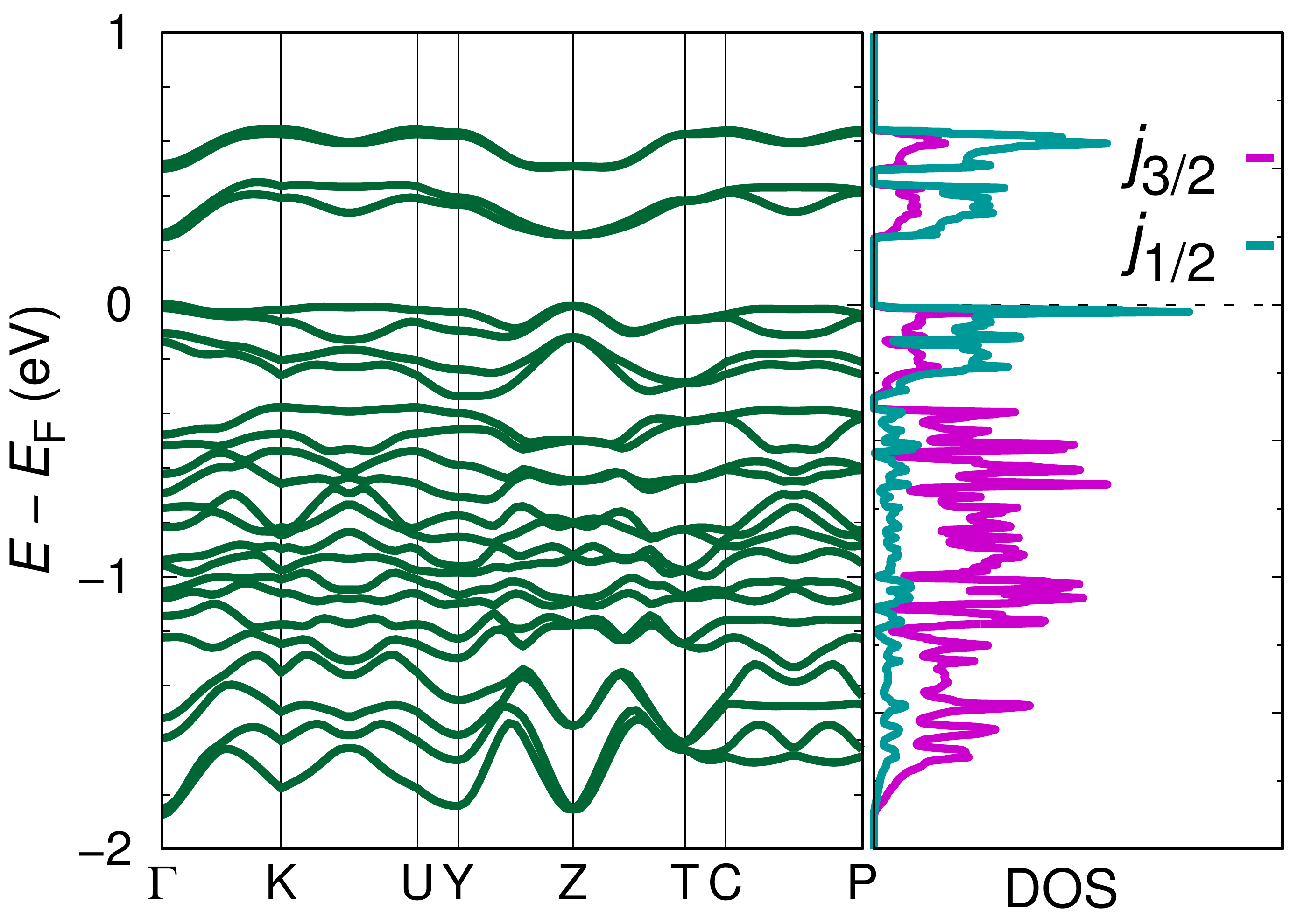}
\caption{(Color online) Ir 5$d$ $t_{2g}$ band structure and
  relativistic DOS for {\gliiro} in zigzag magnetic order, obtained
  with GGA+SO+U (U$_{\rm eff}$ = U-J$_{\rm H}$=2.4~eV).}
\label{fig:banddos-ggasou}
\end{figure}

According to experiment, the magnetic ground state in {\gliiro} is spin
spiral~\cite{Biffin2014gamma} and the magnetic structure shows that
the zigzag chains in the $a$ direction are connected along the
$c$ direction (see Fig.~\ref{fig:structure} (c)). In order to
perform spin-polarized DFT calculations in the magnetically ordered
state, we employed a collinear zigzag magnetic configuration with spin
polarization along the $c$ direction as an approximate
representation of the ordered configuration~\cite{Biffin2014gamma}.
Calculations with the spin polarization along $a$ are shown in
Appendix B for comparison. Inclusion of $U$ within the
GGA+SO+U approach in the zigzag magnetic configuration
(Fig.~\ref{fig:structure}(c)) opens a gap of 242 meV
(Fig.~\ref{fig:banddos-ggasou}) which is smaller than the
experimentally measured value of 0.5 eV~\cite{Hinton2015}.  We note
that the size of the gap is influenced by the choice of $U$. We however
decided here to use the same $U$ parameter as for previous
calculations for {\nairo} and {\aliiro}~\cite{Li2015} in order to
allow a better comparison below.  The magnetic moment
converged to 0.58 ${\mu_{\rm B}}$ for Ir(1) and 0.44 ${\mu_{\rm B}}$
for Ir(2).

\subsection{Exact Diagonalization of Finite Clusters}

While the GGA+SO+U calculations are able to describe many significant
aspects of the electronic structure of {\gliiro} they do not fully
capture effects originating from correlations beyond GGA+SO+U, which
are expected to be relevant when analyzing electronic excitations.
Therefore, we consider here a complementary approach to DFT, namely
exact diagonalization of the fully interacting Hamiltonian on finite
clusters~\cite{BHKim2014} and compare with DFT results.

\begin{figure}
\includegraphics[angle=0,width=0.95\linewidth]{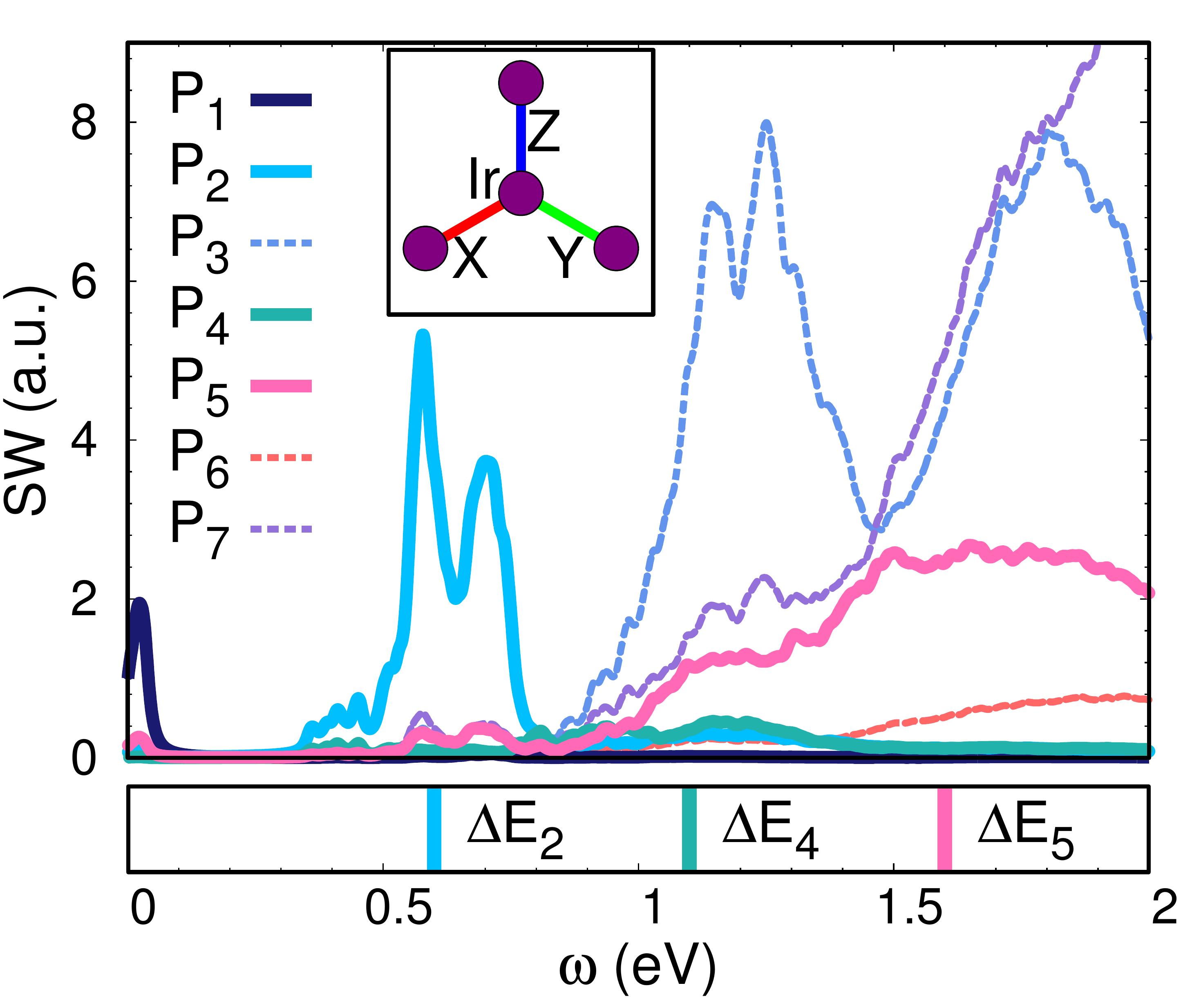}
\caption{(Color online) The investigated four-site cluster (inset) and
  spectral weight (SW) of projected excitations spectra for {\gliiro}.
  P$_1$ includes all states with $(j_{3/2})^4(j_{1/2})^1$
  ($\mathcal{S}_1$), P$_2$ and P$_3$ include the states with an
  exciton on one site ($\mathcal{S}_2$) or on more sites
  ($\mathcal{S}_3$), respectively. P$_4$ includes states from
  $\mathcal{S}_1$ with promotion of an electron $j_{1/2}\rightarrow
  j_{1/2}$ to another site ($\mathcal{S}_4$), and P$_5$ includes
  states with promotion of an electron $j_{3/2}\rightarrow j_{1/2}$ to
  another site ($\mathcal{S}_5$).  P$_6$ is for states that contain
  two-particle excited states for which the $d^4$ site contains
  occupancies $(j_{3/2})^2(j_{1/2})^2$ ($\mathcal{S}_6$), while P$_7$
  includes all other excitations with occupancy of $d^4-d^6-d^5-d^5$
  ($\mathcal{S}_7$). $\Delta E_2 \sim 0.6$ eV, $\Delta E_4 \sim $ 1.1
  eV, $\Delta E_5 \sim 1.6$ eV are the excitation energies for P$_2$,
  P$_4$ and P$_5$ respectively.  }
\label{fig:gliiro-pes}
\end{figure}

We have employed four-site clusters shown in the inset of
Fig.~\ref{fig:gliiro-pes} and obtained the exact eigenstates of the Ir
$t_{2g}$-only Hamiltonian described in Ref.~\onlinecite{Winter2016}:
\begin{eqnarray}
\mathcal{H}_{\rm tot} = \mathcal{H}_{\rm hop}+\mathcal{H}_{\rm CF}+\mathcal{H}_{\rm SO} +
\mathcal{H}_{U}
\end{eqnarray}
including the kinetic hopping term $\mathcal{H}_{\rm hop}$, the
crystal field splitting $\mathcal{H}_{\rm CF}$, spin-orbit coupling
$\mathcal{H}_{\rm SO}$, and Coulomb interaction $\mathcal{H}_{U}$. In terms of the $t_{2g}$ basis introduced above, spin-orbit coupling (SOC) is described by:
\begin{align}
\mathcal{H}_{\rm SO}=\frac{\lambda}{2} \sum_i \vec{\mathbf{c}}_{i}^\dagger\left(\begin{array}{ccc} 0 & -i \sigma_z & i \sigma_y \\ i \sigma_z & 0 & -i\sigma_x \\ -i \sigma_y & i\sigma_x & 0\end{array} \right)\vec{\mathbf{c}}_i
\end{align}
where $\sigma_\mu$, $\mu=\{x,y,z\}$ are Pauli matrices. The Coulomb terms are:
\begin{align}
\mathcal{H}_{U}& \ = U \sum_{i,a} n_{i,a,\uparrow}n_{i,a,\downarrow} + (U^\prime - J_{\rm H})\sum_{i,a< b, \sigma}n_{i,a,\sigma}n_{i,b,\sigma} \nonumber \\
 &+ U^\prime\sum_{i,a\neq b}n_{i,a,\uparrow}n_{i,b,\downarrow} - J_{\rm H} \sum_{i,a\neq b} c_{i,a\uparrow}^\dagger c_{i,a\downarrow} c_{i,b\downarrow}^\dagger c_{i,b\uparrow}\nonumber \\ & + J_{\rm H} \sum_{i,a\neq b}c_{i,a\uparrow}^\dagger c_{i,a\downarrow}^\dagger c_{i,b\downarrow}c_{i,b\uparrow} 
\end{align}
where $c_{i,a}^\dagger$ creates a hole in orbital $a\in\{d_{yz},d_{xz},d_{xy}\}$ at site $i$; $J_{\rm H}$ gives the strength of Hund's coupling, $U$ is the {\it intra}orbital Coulomb repulsion, and $U^\prime=U-2J_{\rm H}$ is the {\it inter}orbital repulsion.  For $5d$ Ir$^{4+}$, we take $U=1.7$ eV, $J_{\rm H}=0.3$ eV\cite{Imada2014}. For the four-site clusters, we retain all hoppings including second neighbour.


\begin{figure}
\includegraphics[angle=0,width=0.5\textwidth]{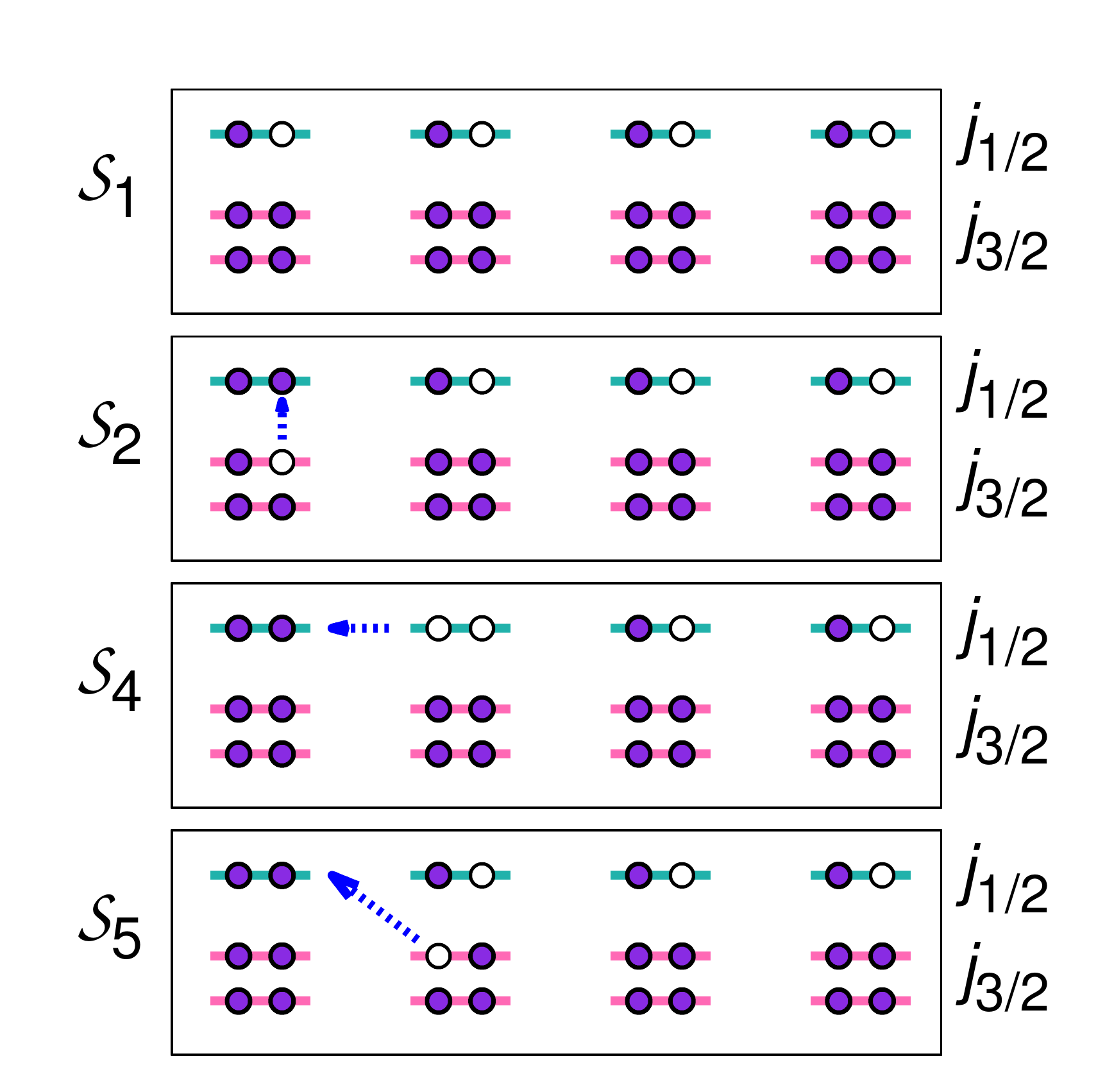}
\caption{(Color online) Schematic diagrams of lowest energy subspace $\mathcal{S}_1$ and one particle excitations
  $\mathcal{S}_2$, $\mathcal{S}_4$, and $\mathcal{S}_5$. Solid circles
  indicate electrons while empty circles are holes. $\mathcal{S}_1$
  are all the states with $(j_{3/2})^4(j_{1/2})^1$, $\mathcal{S}_2$
  are the states obtained from $\mathcal{S}_1$ by promoting an
  electron $j_{3/2}\rightarrow j_{1/2}$ on the same
  site. $\mathcal{S}_4$ are the states obtained from $\mathcal{S}_1$
  by promoting an {\it intersite} $j_{1/2}\rightarrow j_{1/2}$
  transition, and $\mathcal{S}_5$ are the states with promotion of an
  electron $j_{3/2}\rightarrow j_{1/2}$.}
\label{fig:subspace}
\end{figure}

For
{\gliiro}, there are four translationally inequivalent clusters constructed
from bonds ($X_A$, $Y_A$, $Z_A$), ($X_B$, $Y_B$, $Z_B$), ($X_A$,
$Y_A$, $Z_C$), and ($X_B$, $Y_B$, $Z_C$). Of these, the first two are related to one another by 2-fold rotation and the last two are also related by 2-fold rotation. The results presented
correspond to an average over these four clusters. In each four-site cluster, we consider states with a total of four
holes in the $t_{2g}$ orbitals; each Ir site contains six relativistic
orbitals including two $j_{\rm eff}=1/2$ and four $j_{\rm eff}=3/2$
levels. As in Ref.~\onlinecite{BHKim2014}, the many-body basis states for the cluster
can be divided into several subspaces based on the occupancy of the
various orbitals and sites.  Basis states with site occupancy
$d^5-d^5-d^5-d^5$ are included in subspaces
$\mathcal{S}_1-\mathcal{S}_3$, states with site occupancy
$d^4-d^6-d^5-d^5$ belong to $\mathcal{S}_4-\mathcal{S}_7$, and
$\mathcal{S}_8$ contains all higher excitations. We show representative diagrams
of the lowest energy subspace $\mathcal{S}_1$ and one
particle excitation $\mathcal{S}_2$, $\mathcal{S}_4$, and
$\mathcal{S}_5$ in Fig.~\ref{fig:subspace}. Subspace $\mathcal{S}_1$
contains all states with $(j_{3/2})^4(j_{1/2})^1$ occupancy at every
site, which represent a significant contribution to the ground state and
low-lying magnon-like spin excitations.

From these configurations, promotion of a single electron via {\it onsite}
$j_{3/2}\rightarrow j_{1/2}$ generates subspace $\mathcal{S}_2$, containing all
states with a single spin-orbital exciton; the characteristic
excitation energy for such states is given by $\Delta E_2 \sim 3\lambda/2 \sim 0.6$
eV. All states with multiple excitons are grouped into subspace $\mathcal{S}_3$,
and represent $n$-particle excitations from the ground state, with
energies $\Delta E_3 \sim 3n\lambda/2 \sim$ 1.2, 1.8, ... eV. 

Starting from $\mathcal{S}_1$, promotion of an electron via {\it
  intersite} $j_{1/2}\rightarrow j_{1/2}$ yields subspace $\mathcal{S}_4$, containing states with
characteristic energy $\Delta E_4 \sim \mathbb{A}^{-1}$,
where\cite{Winter2016}:
\begin{align}
\mathbb{A} = & \ -\frac{1}{3}\left\{\frac{J_H + 3(U+3\lambda)}{6J_H^2 - U(U+3\lambda)+J_H(U+4\lambda)} \right\}
\end{align}
Taking $U=1.7$ eV, $J_H = 0.3$ eV, and $\lambda = 0.4$ eV suggests $\Delta E_4 \sim $ 1.1 eV.

Starting from $\mathcal{S}_1$, promotion of an electron via {\it intersite}  $j_{3/2}\rightarrow j_{1/2}$
 yields subspace $\mathcal{S}_5$, containing states with characteristic energy $\Delta E_5 \sim \mathbb{C}^{-1} \sim 1.6$ eV, where\cite{Winter2016}:
\begin{align}
\mathbb{C} =& \  \frac{6}{8}\left\{ \frac{1}{2U-6J_H+3\lambda}+\frac{5}{9}\frac{(3U-7J_H+9\lambda)}{J_H}\eta\right\}\\
\eta= & \ \frac{J_H}{6J_H^2-J_H(8U+17\lambda)+(2U+3\lambda)(U+3\lambda)}
\end{align}
Subspace $\mathcal{S}_6$ contains two-particle excited states for
which the $d^4$ site contains occupancies $(j_{3/2})^2(j_{1/2})^2$,
while subspace $\mathcal{S}_7$ contains all other excitations with
occupancy of $d^4-d^6-d^5-d^5$. Single particle excitations most
relevant for the optical conductivity in the next section are contained in $\mathcal{S}_1,
\mathcal{S}_4, \mathcal{S}_5$. The effect of intersite hopping (and Hund's coupling) is to mix states from different subspaces, but the characteristic energies remain valid.

 In order to show this, we project the exact cluster eigenstates $\phi_m$
 on different subspaces:
\begin{align}
\Gamma^m_i = \sum_{s \in \mathbf{\mathcal{S}_i}} \left | \left\langle \phi_m | s \right\rangle \right |^2, 
\label{eq:pescof}
\end{align}
and take the spectral weight (SW) of the projected excitation spectra $P_i$~\cite{BHKim2012}:
\begin{align}
P_i \left( \omega \right ) =\sum_m \Gamma^m_i \delta \left(\omega - E_m\right).
\label{eq:pes}
\end{align}
$P_1$ to $P_7$ are shown in Fig.~\ref{fig:gliiro-pes}. As expected,
the ground state and low-lying magnon-like spin excitations ($\omega
\sim 0$ eV) have the dominant $\mathcal{S}_1$ character (large $P_1$),
while intersite hopping perturbatively mixes in some $\mathcal{S}_2$,
$\mathcal{S}_4$, $\mathcal{S}_5$ character. Indeed, from the localized picture, it is the {\it intersite}
$j_{3/2}\rightarrow j_{1/2}$ mixing that is the
origin of the anisotropic Kitaev exchange couplings.

Regarding the higher excitations: centered at $\omega =
\Delta E_2 \sim 0.6$ eV are the single exciton-like states, with
dominant $\mathcal{S}_2$ character. These states weakly mix with the
single-particle $\mathcal{S}_4$ and $\mathcal{S}_5$ and multi-particle
$\mathcal{S}_6$ and $\mathcal{S}_7$ excitations via intersite
hopping. As expected, excitations with dominant $\mathcal{S}_4$
character (i.e. $j_{1/2}\rightarrow j_{1/2}$) are centered around
$\omega = \Delta E_4 \sim 1.1$ eV, and excitations with dominant
$\mathcal{S}_5$ character (i.e. $j_{3/2}\rightarrow j_{1/2}$) are
centered around $\omega = \Delta E_5 \sim 1.6$ eV. The widths of these
bands are approximately 1 eV and 2 eV, respectively, which is consistent with the GGA+SO+U results above. It is worth
noting that the total spectral weight $\int P_i \ d\omega$ is much
larger for $\mathcal{S}_5$ than $\mathcal{S}_4$, such that
$j_{3/2}\rightarrow j_{1/2}$ excitations dominate the projected
excitation spectra. Similar results were obtained in
Ref.~\onlinecite{BHKim2014} in the analysis of the excitation spectrum
of {\nairo}.

\section{Optical Conductivity \label{sec-D}}

We employ two methods to compute the optical conductivity. The interband contribution to the real part of the optical
conductivity in the DFT calculations is approximated
by\cite{Draxl2006,Ferber2010}:
\begin{equation}\begin{split}
    \sigma_{\mu \nu }(\omega )\propto\frac{1}{\omega}\sum_{c,v}\int d{\bf k}&\left\langle c_{\bf k}|p^{\mu }|v_{\bf k}\right\rangle\left\langle v_{\bf k}|p^{\nu }|c_{\bf k}\right\rangle\\
    &\times\delta (\varepsilon _{c_{\bf k}}-\varepsilon _{v_{\bf
        k}}-\omega ).
\end{split}
\label{eq:op-dft}
\end{equation}
where $\mu$ and $\nu$ correspond to the cartesian axes $x^\prime$,
$y^\prime$, $z^\prime$, which is chosen as the direction of
$a$, $b$, $c$ in this system. $\omega$
indicates the energy of the incident photon, and $p$ is the momentum
operator. The states $|c_{\bf k}\rangle$ in $\mathbf{k}$ space with energy
$\varepsilon _{c_{\bf k}}$ belong to occupied single-particle states,
while $|v_{\bf k}\rangle$, $\varepsilon _{v_{\bf k}}$ describe unoccupied
states.

For the exact diagonalization calculations, we calculate the real part of the optical conductivity at finite temperature
using\cite{BHKim2012}
\begin{align}
\sigma_{\mu\nu}(\omega ) \propto &\frac{\pi (1-e^{-\omega/(k_BT)})}{\omega V}\sum_{nm}B_nM_{\mu,\nu}^{m,n}\delta(\omega +E_{n}-E_{m})
\label{eq:op-ed}
\end{align}
where V is the volume, $B_n$ is the probability density of eigenstate
$\left |\phi_n \right\rangle $.
\begin{align}
B_n =\frac{e^{-\beta E_{n}}}{\sum_{n^\prime} e^{-\beta E_{n^\prime}}}
\label{eq:Bn}
\end{align}
and $M_{\mu,\nu}^{m,n}$ are matrix elements of the current operator:
\begin{align}
M_{\mu,\nu}^{m,n} =\left\langle n\left|j_{\mu }\right|
m\right\rangle \left\langle m\left|j_{\nu }\right| n\right\rangle
\label{eq:M}
\end{align}
The current operator $j_{\mu}$ is given by\cite{Baeriswyl1987}:
\begin{equation}
 j_{\mu }=\frac{ie}{\hbar}\sum_{\begin{subarray}{c}i<j\\ a,b,\sigma,\sigma^\prime\end{subarray}}(c_{i,a,\sigma}^{\dagger}c_{j,b,\sigma^\prime}-c_{j,b,\sigma^\prime}^{\dagger}c_{i,a,\sigma})\ t^{a,b}_{i,j}\mathbf{r}_{ij}^\mu,
\label{eq:jop}
\end{equation}
where $t^{a,b}_{i,j}$ are the hopping parameters between the
$t_{2g}$ orbitals and $\mathbf{r}_{ij}^\mu$ is the $\mu$ component of
the vector from site $j$ to site $i$.  Note that the expression of the
optical conductivity considered in Eq.~(\ref{eq:op-dft}) is defined at
zero temperature and in $k$ space while in Eq.~(\ref{eq:op-ed}) we consider
the definition in real space and at finite temperature $k_{\rm B}T$ = 30
meV (room temperature). We observe that the finite temperature
modifies the zero temperature results only slightly.  The optical
conductivity is normalized by the sum-rule that the energy integral of
the optical conductivity in both ED and DFT methods is proportional to
the effective density of electrons.
\begin{figure}
\includegraphics[width=0.90\linewidth]{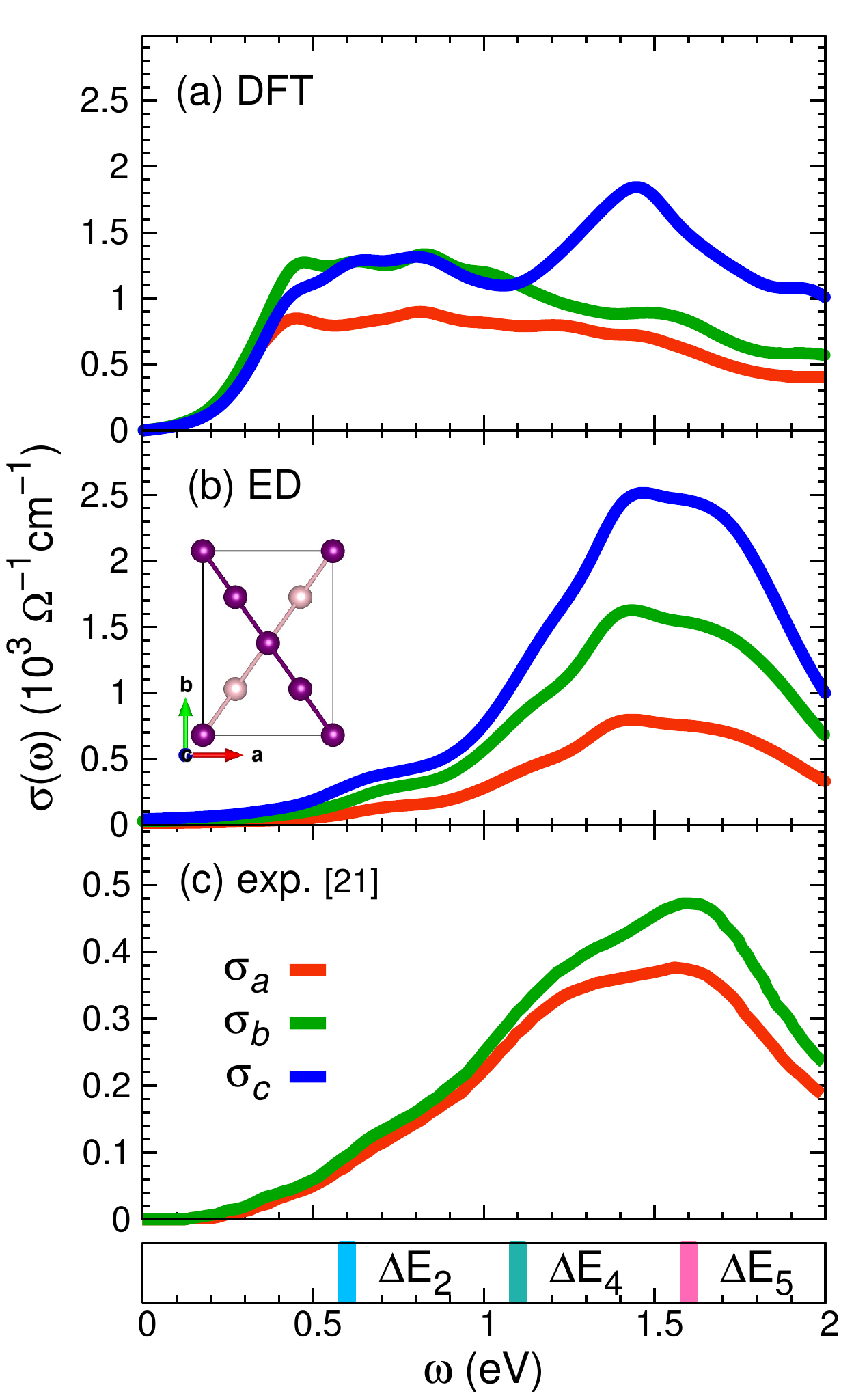}
\caption{(Color online) Optical conductivity components for {\gliiro}.
  (a) Results from DFT within GGA+U+SO. (b) Results from exact
  diagonalization and (c) results reported from experimental
  observations~\cite{Hinton2015}. $\Delta E_2 \sim 0.6$ eV, $\Delta
  E_4 \sim $ 1.1 eV, $\Delta E_5 \sim 1.6$ eV are the characteristic excitation
  energy for subspaces $\mathcal{S}_2$, $\mathcal{S}_4$ and $\mathcal{S}_5$ respectively.  The inset of (b) is
  the crystal structure projected in the $ab$-plane.  }
\label{fig:g-optabc}
\end{figure}

For {\gliiro}, the orthorhombic symmetry of the space group allows the
optical conductivity tensor to be defined in terms of the three
independent components $\sigma _{a}$, $\sigma _{b}$, $\sigma _{c}$
($\sigma_{a}$= $\sigma_{x^\prime x^\prime}$, $\sigma_{b}$=
$\sigma_{y^\prime y^\prime}$, $\sigma_{c}$= $\sigma_{z^\prime
  z^\prime}$):

\begin{figure}
\includegraphics[angle=0,width=0.95\linewidth]{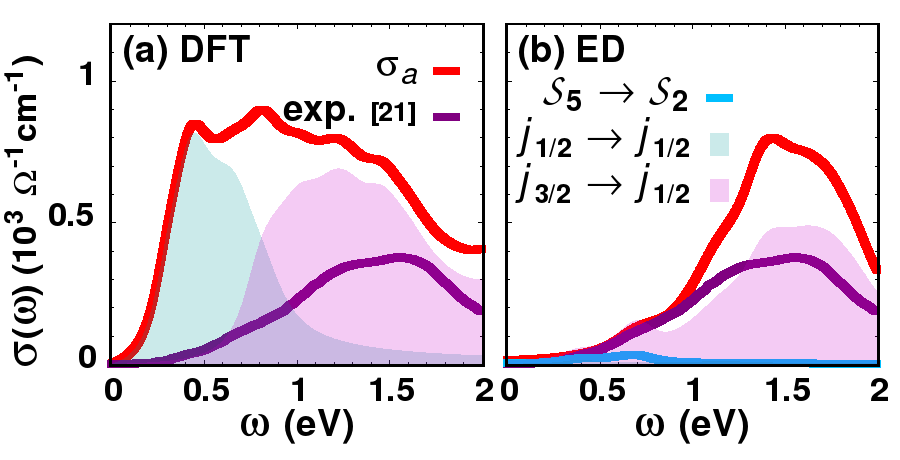}
\caption{(Color online) Optical conductivity component $\sigma_a$ for
  {\gliiro} and different $d$-$d$ transitions in the relativistic basis
  calculated (a) with DFT within GGA+U+SO and (b) with the ED
  method. The comparison to experiment is also
  shown~\cite{Hinton2015}.}
\label{fig:op-grel}
\end{figure}

\begin{equation}
\begin{pmatrix} J_a\\ J_b\\ J_c \end{pmatrix}
=\begin{pmatrix} {\sigma}_{a}&0&0\\0&{\sigma}_{b}&0\\0&0&{\sigma_{c}} \end{pmatrix}
\begin{pmatrix} E_a\\ E_b\\ E_c \end{pmatrix}.
\end{equation}
In Fig.~\ref{fig:g-optabc}, we compare the DFT (GGA+U+SO), ED and
experimental optical conductivity tensor components for {\gliiro}. Both DFT and ED capture correctly the
anisotropy $\sigma_a$ $<$ $\sigma_ b$ $<$ $\sigma_ c$, which is due to the structural
orientation of the planes shown in the inset of
Fig.~\ref{fig:g-optabc} (b). Given that interplane hopping is very weak, the in-plane component of $\sigma(\omega)$ dominates. The magnitude of $\sigma(\omega)$ for polarization along each axis is therefore related to the projection of that axis on to the Ir planes. For light polarized along the $c$-axis, the response is solely due to in-plane processes, while polarization along the $a$- or $b$-axes measures only a fraction of the in-plane response. This observation explains the reduction of the measured $\sigma(\omega)$ $(\sigma_a, \sigma_b)$ for $\gamma$-Li$_2$IrO$_3$ discussed in Ref. \onlinecite{Hinton2015}, when compared with the in-plane measurements of Na$_2$IrO$_3$.

While ED calculations show a dominant
peak around $\omega=$ 1.6~eV for all polarizations, consistent with the experimental data, the DFT results
suggest also significant spectral weight at lower frequencies. The origin of this anomalous
spectral weight can be found in Fig.~\ref{fig:op-grel}. For the DFT
calculations, we show the decomposition of $\sigma(\omega)$ into
intraband $j_{1/2}\rightarrow j_{1/2}$ and interband
$j_{3/2}\rightarrow j_{1/2}$ excitations. For the ED
calculations, we plot the projection of $\sigma(\omega)$ onto
the $\mathcal{S}_1\rightarrow \mathcal{S}_2$ (i.e. spin-orbital excitons), $\mathcal{S}_1 \rightarrow
\mathcal{S}_4$ (i.e. $j_{1/2}\rightarrow j_{1/2}$), and $\mathcal{S}_1
\rightarrow \mathcal{S}_5$ (i.e. $j_{3/2}\rightarrow j_{1/2}$)
excitations. Although direct $\mathcal{S}_1\rightarrow \mathcal{S}_2$ excitations are optically forbidden, the spin-orbital excitonic states $\mathcal{S}_2$ also make a weak contribution to $\sigma(\omega)$ in the mid-energy range due to weak higher order effects. These contributions are also shown. Both the DFT and ED calculations suggest that the peak around 1.6~eV is
due primarily to interband $j_{3/2}\rightarrow j_{1/2}$
contributions. The anomalous low-frequency ($\omega < 1$ eV) spectral weight in the DFT arises primarily from $j_{1/2}\rightarrow j_{1/2}$ excitations between the upper and lower Hubbard bands, the intensity of which are dramatically suppressed in the ED results. This difference can be traced back to two main effects: 

(i) From a localized perspective, we can consider the ground state for two sites as having an electronic configuration $\mathcal{S}_1$ : site 1 = $(j_{3/2})^4(j_{1/2})^1$, site 2  = $(j_{3/2})^4(j_{1/2})^1$. Intersite $j_{1/2}\rightarrow j_{1/2}$ transitions yield local configurations like $\mathcal{S}_4 : (j_{3/2})^4(j_{1/2})^0-(j_{3/2})^4(j_{1/2})^2$, which have a low spin degeneracy as a result of the filled or empty $j_{1/2}$ states. Intersite $j_{3/2}\rightarrow j_{1/2}$ excitations yield local configurations like $\mathcal{S}_5 : (j_{3/2})^3(j_{1/2})^1-(j_{3/2})^4(j_{1/2})^2$, which have a larger spin-degeneracy due to the partially filled $j_{3/2}$ and $j_{1/2}$ states. Overall, the ratio of the total spectral weight associated with these transitions should be $\int P_4 (\omega):\int P_5 (\omega) = 1:8$, as shown in Fig. \ref{fig:gliiro-pes}. In contrast, the DFT calculations take an effective single-particle momentum space perspective, in which the $j_{3/2}$ band is fully occupied, and the $j_{1/2}$ band is half-occupied. The spectral weight associated with $j_{1/2}\rightarrow j_{1/2}$ and $j_{3/2}\rightarrow j_{1/2}$ transitions is therefore $1:4$, which overestimates the contributions of the former in DFT calculations compared to the localized picture. In other words, DFT does not correctly capture the spin-multiplicity associated with the localized states. 

(ii) In a localized picture, the current operator depends on the intersite
hopping matrix elements via Eq. (\ref{eq:jop}). It is therefore useful to
rewrite the nearest neighbour hopping integrals in the relativistic basis. For
example, for the $Z$-bonds, these are:
\begin{align}
t_{ij}(j_{1/2}\rightarrow j_{1/2}) & \propto  (2t_1 + t_3) \\
t_{ij}(j_{3/2}; m_{\pm 1/2} \rightarrow j_{1/2}) & \propto (t_3-t_1)\\
t_{ij}(j_{3/2}; m_{\pm 3/2} \rightarrow j_{1/2}) & \propto t_2
\end{align}
Via the current operator Eqs. (15)-(16), the optical conductivity associated
with each transition scales with $\sigma(\omega) \propto (t_{ij})^2$. Typically,
in the corner-sharing iridates such as $\gamma$-Li$_2$IrO$_3$, $t_1$ and $t_3$
have opposite sign (and may be quite small), which suppresses the
$(j_{1/2}\rightarrow j_{1/2})$ hopping, reducing the influence of such
excitations on the optical conductivity. This effect is partially captured in
DFT, as can be seen from comparing the relative widths of the $j_{1/2}$ and
$j_{3/2}$ bands in Fig. \ref{fig:banddos-ggasou}. However, DFT likely
overestimates the degree of $j_{1/2}-j_{3/2}$ mixing which leads, effectively,
to larger optical matrix elements between low-energy states.

Overall, we conclude that the ED calculations, based on DFT hopping integrals, provides the best description of the optical conductivity.

\section{Comparison to {\protect\nairo} and {\protect\aliiro}\label{sec-3}}

\begin{figure}
\includegraphics[angle=0,width=0.95\linewidth]{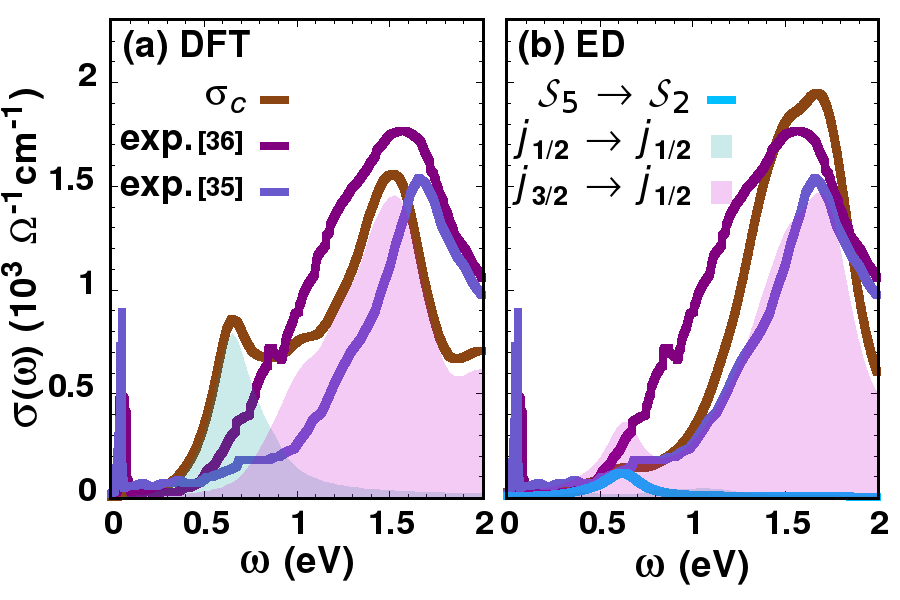}
\caption{(Color online) Optical conductivity component $\sigma_{c}$ for {\nairo}  and different $d$-$d$ transitions in the
relativistic basis calculated (a) with DFT within GGA+U+SO and (b) with the ED method.  Comparison with experimental
results from Ref.~\onlinecite{Comin2012} and Ref.~\onlinecite{Sohn2013} is also shown.
 $\sigma_{c}$ of {\nairo} corresponds to the $\sigma_{zz}$ component in Ref.~\onlinecite{Li2015}.}
\label{fig:nairo-oprel}
\end{figure}

Despite differences in crystal architecture, the experimental optical
conductivity of $\gamma$-Li$_2$IrO$_3$ and {\nairo} share a
very similar profile that we will analyze in what follows.  As stated
in the previous section, $\sigma(\omega)$ should be
dominated by intersite $j_{3/2}\rightarrow j_{1/2}$ excitations, at
$\omega \sim \mathbb{C}^{-1} \sim 1.6$ eV, as observed. The soft
shoulder observed at lower energies results from a combination of low
spectral weight from intersite $j_{1/2}\rightarrow j_{1/2}$
excitations centered at $\omega \sim \mathbb{A}^{-1} \sim 1.1$ eV, and
weak mixing with optically forbidden local $j_{3/2}\rightarrow
j_{1/2}$ excitons near $\omega \sim $ 0.6 eV. These assignments are
consistent with the fitting of $\sigma (\omega)$ in
Ref.~\onlinecite{Sohn2013} for {\nairo}, which suggested peaks in the
vicinity of 0.72, 1.32, and 1.66 eV. However, the ``band gap''
reported to be 0.32 eV is likely to be significantly contaminated by low-lying
excitonic states, and may therefore not represent the natural charge
gap of the material. The origin of the peaks for {\nairo} in the relativistic basis are shown in Fig.~\ref{fig:nairo-oprel} for both calculations.

\begin{figure}
\includegraphics[angle=0,width=0.95\linewidth]{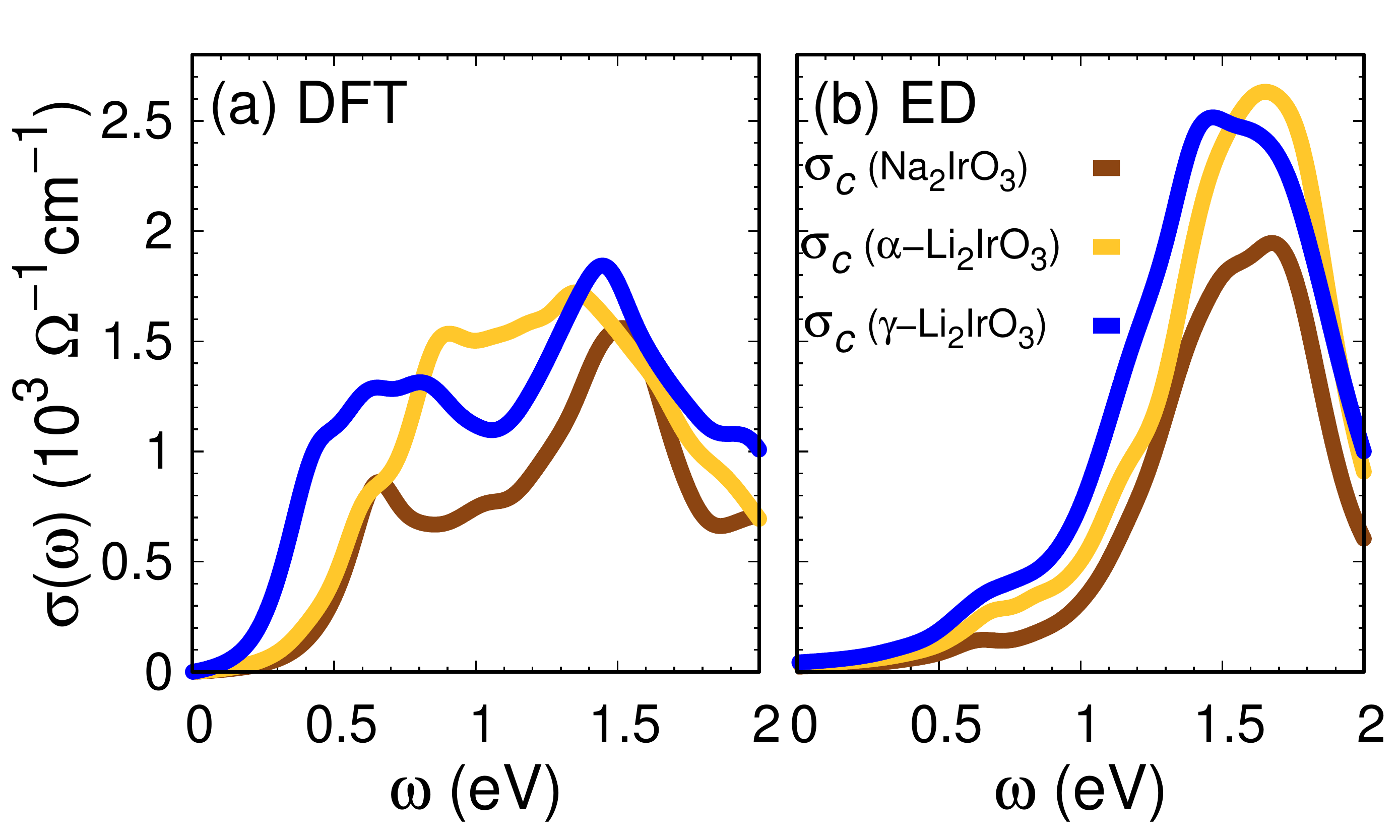}
\caption{(Color online) Optical conductivity $\sigma_{c}$ for {\nairo}, {\aliiro}, and {\gliiro}
calculated (a) with DFT within GGA+U+SO  and (b) with the ED method. $\sigma_{c}$ for {\nairo} and {\aliiro} corresponds
 to  $\sigma_{zz}$  in Ref.~\onlinecite{Li2015}. Please note that
 for {\aliiro} differences in the DFT optical conductivity with respect to
results in Ref.~\onlinecite{Li2015} lie in the employed crystal structure.}
\label{fig:op-compare}
\end{figure}

In Fig.~\ref{fig:op-compare} we display the theoretical DFT and ED results for the
in-plane $\sigma_c$ component for {\nairo}, {\aliiro} and {\gliiro}. For {\aliiro}, we employed the recently obtained single crystal structure~\cite{Freund2016}. Hopping integrals and crystal field parameters for the revised structure are given in the Appendix. It should be noted that the results obtained for $\alpha$-Li$_2$IrO$_3$ in this work therefore differ slightly from the results in Ref.~\onlinecite{Li2015}, which employed instead previously available structures obtained from powder x-ray analysis and structural relaxation. Generally for these materials, the electronic structure is strongly affected by the competition between spin-orbit coupling ($\lambda \sim 0.4$ eV) and crystal-field terms ($\Delta_n \sim 0-0.2$ eV), which leads to an enhanced dependence of the spectra on structural details. Nonetheless, both DFT and ED calculations give a strong main peak in $\sigma(\omega)$ near $\omega=$ 1.6 eV for {\aliiro}, {\gliiro} and {\nairo}.  This peak is predicted to be more intense in the former two materials by both methods. Further, both DFT and ED calculations
show an enhanced spectral weight at lower energies in {\gliiro} with
respect to {\nairo}, which is consistent with experimental results.  The differences can be understood as follows. For materials dominated by oxygen-assisted hopping such as {\nairo}, the hopping integrals in the $d$-orbital basis satisfy $t_2 \gg t_1,t_3$, so in the relativistic basis
the hopping is dominated by $t_{ij}(j_{3/2}; m_{\pm 3/2} \rightarrow
j_{1/2})$. This observation suggests negligible spectral weight for $j_{1/2}\rightarrow
j_{1/2}$ excitations in $\sigma(\omega)$. In contrast, for significant direct metal-metal hopping
$t_1$, $t_3$, additional spectral weight may appear in the mid-energy region due
to enhanced $t_{ij}(j_{1/2}\rightarrow j_{1/2})$. This is worth noting because the values of these hopping integrals are directly related to the magnetic interactions. Indeed, up to second order in hopping,  the magnetic interactions are given by:\cite{Winter2016}
\begin{align}
J_1 =& \ \frac{4\mathbb{A}}{9}(2t_1+t_3)^2-\frac{8\mathbb{B}}{9}\left\{2(t_1-t_3)^2\right\} \label{eqn-J1}\\
K_1 =& \frac{8\mathbb{B}}{3}\left\{ (t_1-t_3)^2-3t_2^2\right\}\label{eqn-K1} \\
\Gamma_1 =& \ \frac{8\mathbb{B}}{3}\left\{2t_2(t_1-t_3)\right\}\label{eqn-G1}
\end{align}
where $\mathbb{B}$ is a constant similar to $\mathbb{A}$ and $\mathbb{C}$:
\begin{align}
\mathbb{B} = & \ \frac{4}{3}\left\{\frac{(3J_{\rm H}-U-3\lambda)}{(6J_{\rm H}-2U-3\lambda)}\eta\right\}
\end{align}
The desirable Kitaev limit ($K_1 \gg J_1,\Gamma_1$) is obtained only for $t_2\gg t_1,t_3$, and will therefore be most closely approached by materials with the low spectral weight
near $\omega \sim 1.1$ eV. This identifies {\nairo} as the closest material to the Kitaev limit from
all three investigated here, in
agreement with Ref.~\onlinecite{Winter2016}.

\section{Summary}
We have investigated the electronic structure, hopping parameters and
optical excitation spectrum of the three-dimensional {\gliiro}.  Due
to the lower symmetry of the local Ir-O-Ir environment, the hopping
integrals display significant deviations from the ideal case,
suggesting e.g. large metal-metal hoppings and departures from
inversion symmetric values. This situation likely leads to highly
complex magnetic interactions in this system and manifests in certain 
signatures in the optical conductivity.

We computed the optical conductivity by two methods; (i) relativistic
DFT calculations within GGA+SO+U and (ii) exact diagonalization of the
full interacting Hamiltonian on finite clusters where the hopping
integrals were obtained from DFT. Both methods reproduce the main peak
of the in-plane component of the optical conductivity
$\sigma_c$. However, GGA+SO+U seems to overestimate the contribution
of the $j_{1/2} \rightarrow j_{1/2}$ transition at low energies in
$\sigma_a$ and $\sigma_b$.  The ED results, in contrast, validate the
model parameters ($U, J_H, \lambda$) and suggest that the high-lying
excitations appear to be well captured within a localized picture in {\gliiro}.
The comparison with the optical conductivity analysis of {\nairo}
shows that the peak near 1.5 eV in both {\nairo} and {\gliiro} can be
identified in terms of {\it intersite} $j_{3/2}\rightarrow j_{1/2}$
excitations.  The comparison of $\sigma(\omega)$ for the various
materials suggests that the relative spectral weight of the
transitions provide insight into the magnitudes of various hopping
integrals, and therefore the local magnetic interactions.

\begin{acknowledgments} We would like to thank Daniel Guterding and
  Kira Riedl for very useful discussions. We would also like to thank Philipp Gegenwart for pointing out the new single crystal structural data of $\alpha$-Li$_2$IrO$_3$. Y.L.  acknowledges support
  through a China Scholarship Council (CSC)
  Fellowship. S. M. W. acknowledges support through an NSERC Canada
  Postdoctoral Fellowship.  H.O.J and R.V. acknowledge support by the
  Deutsche Forschungsgemeinschaft through grant SFB/TR 49.
\end{acknowledgments}

\appendix
\section{Hopping parameters for the nonmagnetic nonrelativistic system}
\begin{figure}
\includegraphics[angle=0,width=0.95\linewidth]{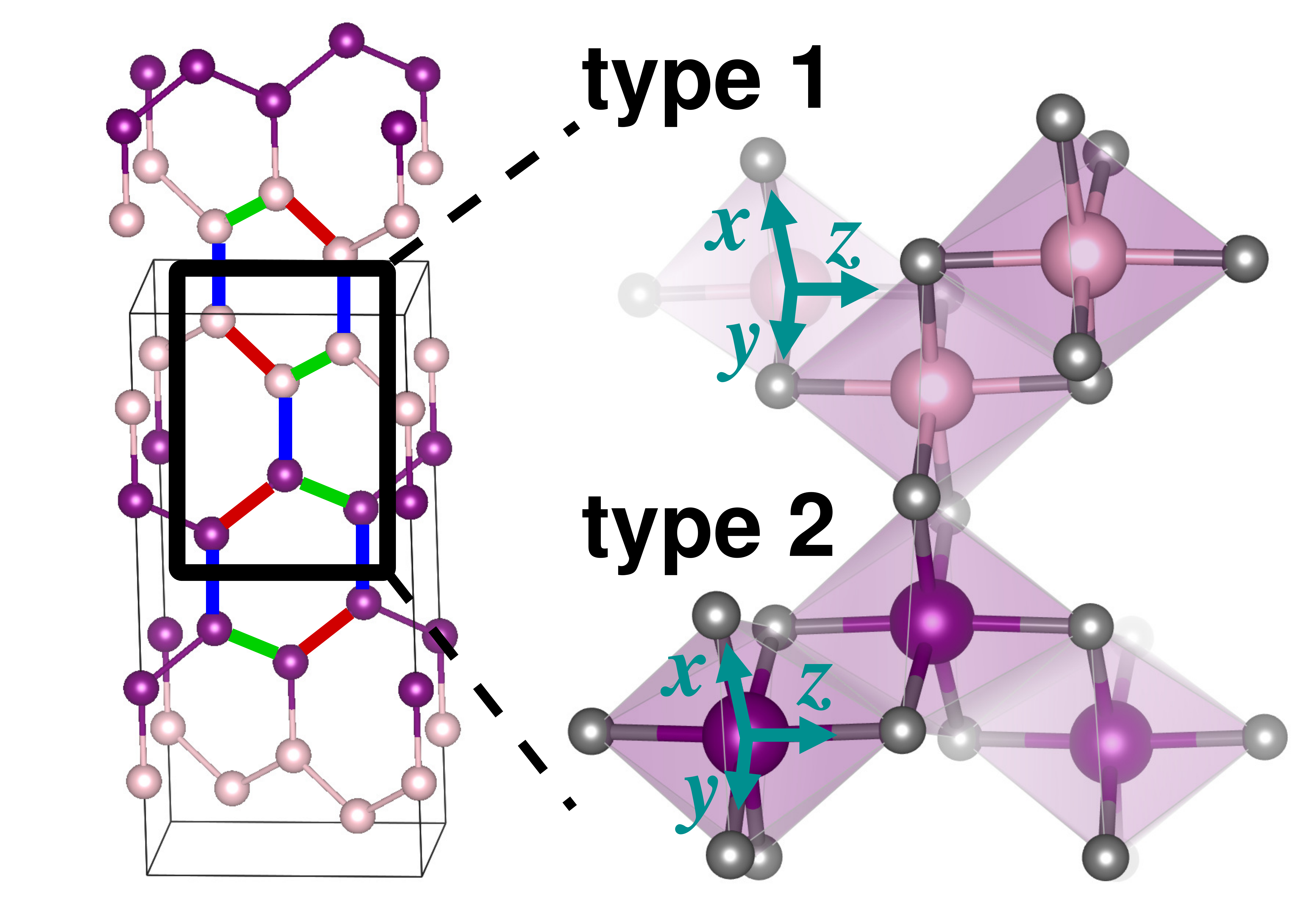}
\caption{(Color online) Local octahedral environment of (a) type 1 and
  (b) type 2 in {\gliiro}. ($n_1$, $n_2$, $n_3$) correspond to
  ($y,z,x$) and ($z,x,y$) for X and Y bonds, respectively.}
\label{fig:hop_config}
\end{figure}

  Table~\ref{table:hopappendix1} and Table~\ref{table:hopappendix2}
show all onsite and nearest neighbor hopping parameters in {\gliiro}. As noted above, the $t_{1O}$ in the $X_A$ ($Y_A$), $X_B$
($Y_B$) bonds have opposite signs, as a result of different local
environments. The negative value corresponds to type 1 bonds in
Fig.~\ref{fig:hop_config}, while the positive values are type 2
bonds in Fig.~\ref{fig:hop_config}. 

Table~\ref{hopappendix3} and Table~\ref{hopappendix4}
show all onsite and nearest neighbor hopping parameters in {\aliiro}.

\begin{table}[ht]
\caption {Hopping parameters for the on-site terms (meV) in {\gliiro}. A is for hexagon including $X_A$, $Y_A$, $Z_A$ bonds while B is for hexagon including $X_B$, $Y_B$, $Z_B$.}
\label{table:hopappendix1}
\centering\def\arraystretch{1.1}
\begin{ruledtabular}
\begin{tabular}{llr}
Ir(1)&  $xy$ $\rightarrow$ $xy$ &-592.6 \\
     &  $xz$ $\rightarrow$ $xz$ & -379.1 \\
     &  $yz$ $\rightarrow$ $yz$ & -379.1 \\
Ir(2)&  $xy$ $\rightarrow$ $xy$ & -651.3 \\
     &  $xz$ $\rightarrow$ $xz$ & -540.4 \\
     &  $yz$ $\rightarrow$ $yz$ & -540.4 \\
Ir(1)& $xy$ $\rightarrow$ $xz$ & 4.2 (A), -4.2 (B) \\
     & $xy$ $\rightarrow$ $yz$ & 4.2 (A), -4.2 (B)\\
     & $xz$ $\rightarrow$ $yz$ & -24.4 \\
Ir(2)& $xy$ $\rightarrow$ $xz$ & 37.4 (A), -37.4 (B)\\
     & $xy$ $\rightarrow$ $yz$ & 37.4 (A), -37.4 (B) \\
     & $xz$ $\rightarrow$ $yz$ & -29.9 \\
\end{tabular}
\end{ruledtabular}
\centering\def\arraystretch{1.1}
\caption{Nearest neighbor tight-binding hopping matrix elements (meV) for  $\gamma-$Li$_2$IrO$_3$.}
\begin{ruledtabular}
\begin{tabular}{lrrrrrrr}
$\gamma-$Li$_2$IrO$_3$ & $X_A$ & $X_B$ &$Y_A$ &$Y_B$& $Z_A$     & $Z_B$ & $Z_C$  \\
\hline
$xy$ $\rightarrow$ $xy$& 91.4  & 91.4  & 91.4  & 91.4  &-319.7 &-319.7  & -17.1 \\
$xz$ $\rightarrow$ $xz$& 69.2  & 69.2  &-168.3 &-168.3 & 91.8  & 91.8   & 77.4  \\
$yz$ $\rightarrow$ $yz$&-168.3 &-168.3 & 69.2  & 69.2  & 91.8  & 91.8   & 77.4  \\
$xy$ $\rightarrow$ $xz$&-262.5 &262.5  &  4.2  &  -4.2 & 63.9  &-63.9   & -18.7 \\
$xz$ $\rightarrow$ $xy$&-240.5 &240.6  & 76.5  & -76.5 & 63.9  &-63.9   & 18.7  \\
$xy$ $\rightarrow$ $yz$&  4.2  &  -4.2 &-262.5 &262.5  & 63.9  &-63.9   & -18.7 \\
$yz$ $\rightarrow$ $xy$& 76.5  & -76.5 &-240.5 &240.6  & 63.9  &-63.9   & 18.7  \\
$xz$ $\rightarrow$ $yz$& -60.2 & -60.2 & -10.6 & -10.6 &132.8  &132.8   & 294.1  \\
$yz$ $\rightarrow$ $xz$& -10.6 & -10.6 & -60.2 & -60.2 &132.8  & 132.8  &294.1  \\
\end{tabular}
\end{ruledtabular}
\label{table:hopappendix2}
\end{table}

\begin{table}[ht]
\caption {Hopping parameters for the on-site terms (meV) for $\alpha$-Li$_2$IrO$_3$ for the recently available single-crystal structure from Ref. \onlinecite{Freund2016}. \label{hopappendix3}}
\centering\def\arraystretch{1.1}
\begin{ruledtabular}
\begin{tabular}{lr}
 $xy$ $\rightarrow$ $xy$ & -401.8 \\
 $xz$ $\rightarrow$ $xz$ & -517.4 \\
 $yz$ $\rightarrow$ $yz$ & -517.4 \\
 $xy$ $\rightarrow$ $xz$ & -39.0 \\
 $xz$ $\rightarrow$ $yz$ & -39.0 \\
 $xz$ $\rightarrow$ $yz$ & -33.5 \\
\end{tabular}
\end{ruledtabular}
\end{table}

\begin{table}[ht]
\caption {Nearest neighbor tight-binding hopping matrix elements (meV) for $\alpha$-Li$_2$IrO$_3$ for the recently available single-crystal structure from Ref. \onlinecite{Freund2016}. \label{hopappendix4}}
\centering\def\arraystretch{1.1}
\begin{ruledtabular}
\begin{tabular}{lrrr}
$\alpha$-Li$_2$IrO$_3$ &X &Y &Z \\
\hline
$xy \rightarrow xy$ &70.2 &70.2 &-139.3 \\
$xz \rightarrow xz$ &83.6 &-124.0 &77.7 \\
$yz \rightarrow yz$ &-124.0& 83.6 &77.7 \\
$xy \rightarrow xz$ &239.0 &-34.9 &-30.7 \\
$xz \rightarrow xy$ &239.0 &-34.9 &-30.7 \\
$xy \rightarrow yz$ &-34.9 &239.0 &-30.7 \\
$yz \rightarrow xy$ &-34.9 &239.0 &-30.7 \\
$xz \rightarrow yz$ &-38.6 &-38.6 &285.5 \\
$yz \rightarrow xz$ &-38.6 &-38.6 &285.5 \\
\end{tabular}
\end{ruledtabular}
\end{table}

\section{Optical conductivity with spin polarized to $a$ direction}
\begin{figure}
\includegraphics[angle=0,width=0.85\linewidth]{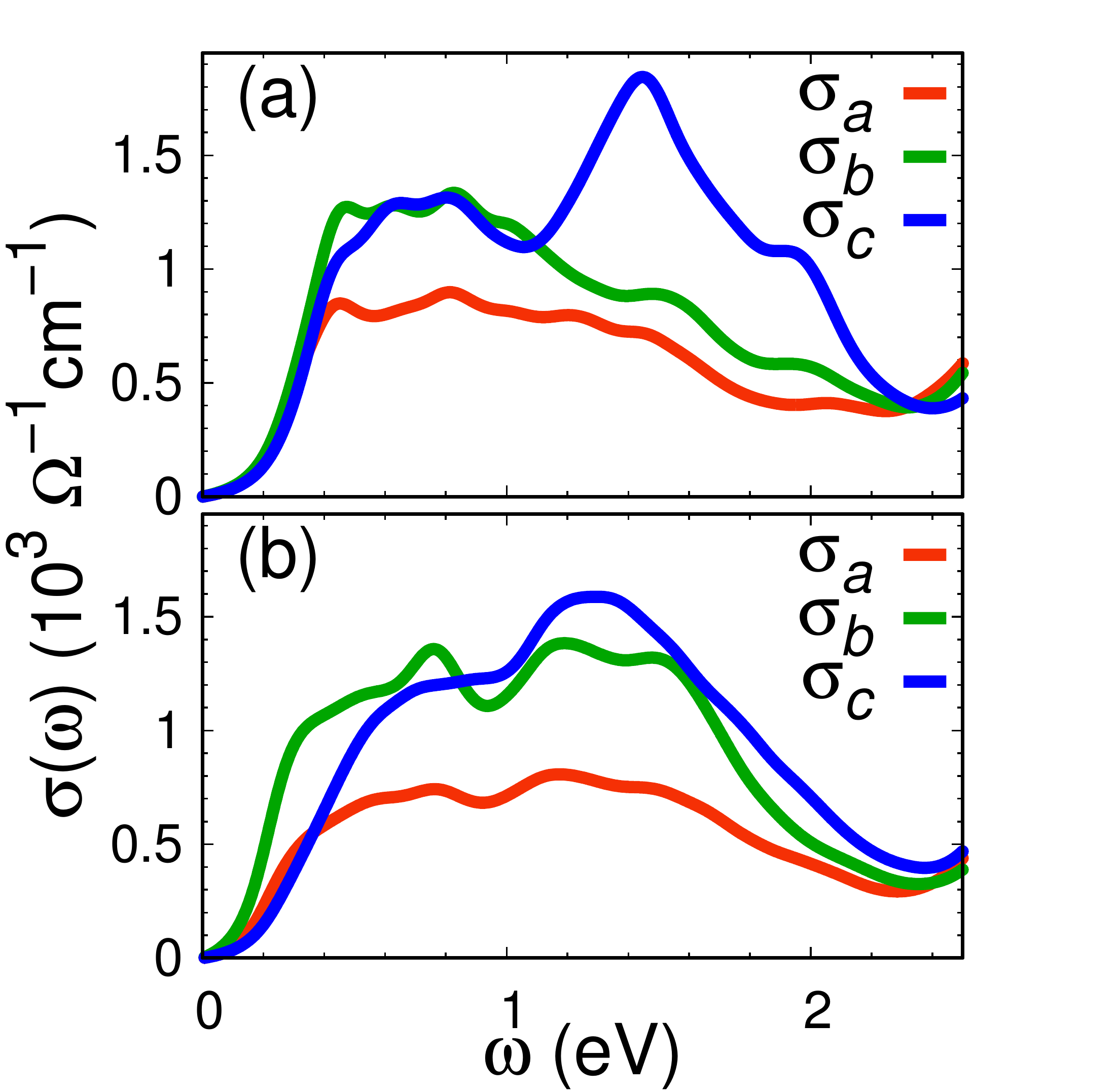}
\caption{(Color online) Optical conductivity tensor components with spin polarization (a) along $c$  and (b) along $a$ in the  zigzag configuration.}
\label{fig:compare}
\end{figure}

In order to compare the dependence of the optical conductivity along various spin  directions in the zigzag magnetic configuration,
 we show the results for spins along $a$ and $c$ direction in Fig.~\ref{fig:compare}. The results
 show that the $\sigma_c$ component doesn't depend significantly on the spin polarized direction, while $\sigma_a$ and $\sigma_b$ are more sensitive to it.

\end{document}